# A New DNA Sequence Vector Space on a Genetic Code Galois Field

Robersy Sánchez [1,2], Ricardo Grau [2,3] and Eberto R. Morgado [3]

[1] Research Institute of Tropical Roots, Tuber Crops and Banana (INIVIT). Biotechnology group. Santo Domingo. Villa Clara. Cuba
{robersy@uclv.edu.cu}

[2] Center of Studies on Informatics, Central University of Las Villas, Villa Clara, Cuba
{rgrau@uclv.edu.cu}

[3] Faculty of Mathematics Physics and Computation, Central University of Las Villas, Villa Clara, Cuba.
(morgado@uclv.edu.cu)

*Abstract—*

A new $N$-dimensional vector space of DNA sequences over the Galois field of the 64 codons ($GF$ (64)) was recently presented. Although in this vector space, gene point mutations were considered linear transformations or translations of the wild type gene, deletions and insertions (indel) could not be considered. Now, in order to include indel mutations, we have defined a new Galois field over the set of elements $X_1X_2X_3$ ($C_{125}$), where $X_i \in$ {O, A, C, G, C}. We have called this set, the extended triplet set and the elements $X_1X_2X_3$, the extended triplets. The order of the bases is derived from the $Z_{64}$-algebra of the genetic code –recently published–. Starting from the natural bijection $\Psi : GF\ (5^3) \to C_{125}$ between the polynomial representation of elements from $GF\ (5^3)$ and the elements $X_1X_2X_3$, a novel Galois field over the set of elements $X_1X_2X_3$ is defined. Taking the polynomial coefficients $a_0, a_1, a_2 \in GF\ (5)$ and the bijective function $f : GF\ (5) \to$ {O, A, C, G, C}, where $f\ (0) =$ O, $f\ (1) =$ A, $f\ (2) =$ C, $f\ (3) =$ G, $f\ (4) =$ U, bijection $\Psi$ is induced such that $\Psi(a_0 + a_1x + a_2x^2) = (f\ (a_1)\ f\ (a_2)\ f\ (a_0)) = (X_1X_2X_3)$. The polynomial coefficient $a_2$ of the terms with a maximal degree $a_2x^2$ corresponds to the base in the second codon position, the coefficient of the term with degree 1 corresponds to the first codon position, and finally, the coefficient of the term of degree 0 is assigned to the third codon position. That is, the degree of the polynomial terms decreases according to the biological meaning of the corresponding base. Next, by means of the bijection $\Psi$ we define sum "+" and product "•" operations in the set of codons $C_{125}$, in such a way that the resultant field ($C_{125}, +, •$) turns isomorphic to the Galois Field $GF\ (5^3)$. In the additive group ($C_{125}, +$), the inverse of codons $X_1AX_3$ that code to hydrophilic amino acids are the codons $(-X_1)U(-X_3)$ which in turn code to hydrophobic amino acids. The sum of a $X_1AX_3$ codon to a $X_1UX_3$ one gives a $X_1OX_3$ codon. Then, this sum introduces at least one base deletion in the extended triplet obtained. The Field ($C_{125}, +, •$) allows the definition of a novel $N$-dimensional vector space ($S$) over the field $GF\ (5^3)$ on the set of all $125^N$ sequences of extended triplets in which all possible DNA sequence alignments of length $N$ are included. Here the "classical gap" produced by alignment algorithms corresponds to the neutral element "O". In the vector space $S$ all mutational events that take place in the molecular evolution process can be described by means of endomorphisms, automorphisms and translations. In particular, the homologous (generalized) recombination between two homologous DNA duplexes involving a reciprocal exchange of DNA sequences –e.g. between two chromosomes that carry the same genetic loci– algebraically corresponds to the action of two automorphism pairs (or two translation pairs) over two paired DNA duplexes. For instance, the automorphism pair $f$ and $f^{-1}$ acts over the homologous DNA strands α and β to turn out the homologous reciprocal recombinants $f(α)$ and $f^{-1}(β)$. Likewise, the pair $g$ and $g^{-1}$ acts over the homologous strands α' and β' to turn out the homologous reciprocal recombinants $g(α')$ and $g^{-1}(β')$.

*Keywords— DNA sequence vector space, Algebraic structure, Gene algebra*

## I. INTRODUCTION

A new $N$-dimensional vector space of DNA sequences over the Galois field of the 64 codons ($GF$ (64)) was recently presented [SAN 05]. This vector space was derived taking into account the order of the bases proposed in the Boolean lattice of the four DNA bases [SAN 04] [SAN 04a]. The isomorphism φ: $B(X) \to (Z_2)^2$ between the Boolean lattices of the four DNA bases $B(X)$ and $((Z^2)^2, \vee, \wedge)$ ($Z^2$={0,1}), and the biological importance of base positions in the codons were used to state a partial order in the codon set. As a result every codon was represented in the field $GF$ (64) as a binary sextuplet.

In this vector space, gene point mutations were considered linear transformations or translations of the wild type gene, however deletions and insertions (indel) could not be considered. Now, in order to include indel mutations, we have defined a new Galois field over the set of elements $X_1X_2X_3$ ($C_{125}$), where $X_i \in$ {O, A, C, G, U}. We call this set, the extended triplet set and the elements $X_1X_2X_3$, the extended triplets. At present, the starting base order used here comes from the recently reported $Z_{64}$-algebra of the genetic code [SAN 05a]. In this $Z_{64}$-algebra the base order {A, C, G, U} was obtained by considering the genetic code as a non-dimensional code scale of amino acid interaction energies in proteins.

Like in previous articles, we have kept in mind the biological importance of base position in the codon to state a codon order in the genetic code. The importance of the base position is suggested by the error frequency (accepted mutations) found in codons. Errors on the third base are more frequent than on the first base, and, in turn, these are more frequent than errors on the second base [WOE 85] [FRI 64] [PAR 89]. These positions, however, are very conservative with respect to changes in polarity of coded amino acids [ALF 69].

The principal aim of this work is to show that all mutational events that take place in the molecular evolution process can be described by means of endomorphisms, automorphisms and translations of a novel $N$-dimensional vector space over the Galois field $GF$ ($5^3$). The new vector space defined over the set of all $125^N$ sequences of extended triplets includes all possible DNA sequence alignments of length $N$. Here the "classical gap" produced by alignment algorithms corresponds to the neutral element "O".

II. THEORETICAL MODEL.

Our starting point is the bases order {A, C, G, U} derived from the $Z_{64}$-algebra of the genetic code [SAN 05a]. In order to analyze indel mutations we extend this alphabet including the new symbol "O" to denote base omissions (gaps) in DNA sequence alignments. As a result, we can build a new triplet set with elements $X_1X_2X_3$ where $X_i \in$ {O, A, C, G, U}. We shall call this set the extended triplet set $C_{125}$ and the elements $X_1X_2X_3$, the extended triplets.

Now, considering the order in the set {O, A, C, G, U} and the biological importance of the base position in the codon, it is possible to establish an order in the extended triplet set, i.e. from triplet OOO to UUU. First, keeping invariables –in the triplets $X_1X_2X_3$– bases $X_1$ and $X_2$, the third base $X_3$ is consecutively changed until all possibilities are exhausted. Next, a similar variation is applied to the first base and finally to the second one, i.e. the variations are introduced from the less biologically relevant base to the most relevant base in the codon. Then, the ordered triplet set showed in the Table 1 was obtained.

## A. Nexus between the Galois Field Elements and the Set of Codons

As one can see in Table 1, a bijection is suggested between the orders in the extended triplet set and the $GF$ ($5^3$) elements. In particular, there is a bijective function $f : GF$ (5) $\rightarrow$ {0, G, U, A, C}, between the elements of $GF$ (5) and the letters $X_k \in$ {O, A, C, G, U}. This function explicitly is given by the equalities:

$$f(0) = O, f(1) = A, f(2) = C, f(3) = G, f(4) = U$$

Next, taking into account the biological importance of base positions in codons, we can state the bijective function $\Psi : GF$ ($5^3$) $\rightarrow C_{125}$ between the extended triplet set and the polynomial representation of $GF$ ($5^3$) elements:

$$\Psi(a_0 + a_1x + a_2x^2) = (f(a_1) f(a_2) f(a_0)) = (X_1X_2 X_3).$$

The polynomial coefficient $a_2$ of the terms with a maximal degree $a_2x^2$ corresponds to the base in the second codon position. The coefficient of the term with degree 1 corresponds to the first codon position, and finally, the coefficient of the term of degree 0 is assigned to the third codon position. That is, the degree of the polynomial terms decreases according to the biological meaning of the corresponding base. Notice that coefficients $a_i$ correspond –for every triplet– to the integer digits of its 3-tuple vector representation in $GF$ ($5^3$).

The reverse of this integer digit sequence corresponds to the integer representation in base 5 of the triplet index number (see Table 1). So, as an example we have the following bijections:

| | | | | | | | |
|---|---|---|---|---|---|---|---|
| 7 | ↔ | 012 | ↔ | 210 | ↔ | $2 + x$ | ↔ | AOC |
| 44 | ↔ | 134 | ↔ | 431 | ↔ | $4 + 3x + x^2$ | ↔ | GAU |
| 117 | ↔ | 432 | ↔ | 234 | ↔ | $2 + 3x + 4x^2$ | ↔ | GUC |

In particular, we will use the bijective function $f[s]$ such that $f: s \rightarrow GF(5^3)$, between the subset of the integer number $s = \{0, 1... 124\}$ and the elements of $GF(5^3)$. According to the above example $f[7] = 2 + x$, $f[44] = 4 + 3x + x^2$ and $f[117] = 2 + 3x + 4x^2$.

## B. Vector Spaces over the Genetic Code Galois Field

Now, by means of the function $\Psi$ we can define a product operation in set $C_{125}$. Let $\Psi^{-1}$ be the inverse function of $\Psi$ then, for all pair of codons $X_1Y_1Z_1 \in C_{125}$ and $X_2Y_2Z_2 \in C_{125}$, their product "•" will be:

$$X_1Y_1Z_1 \bullet X_2Y_2Z_2 = \Psi[\Psi^{-1}(X_1Y_1Z_1) \Psi^{-1}(X_2Y_2Z_2) \bmod g(x)]$$

That is to say, the product between two triplets is obtained from the product of their corresponding polynomial module $g(x)$, where $g(x)$ is an irreducible polynomial of second degree over $GF(5)$. Since there are 40 irreducible polynomials of second degree, we have 40 possible variants to choose the product between two extended triplets. It is not difficult to prove that the set of codons $C_{125}\backslash\{OOO\} = C_{125}^*$ with the operation product "•" is an Abelian group ($C_{125}^*$, •). Likewise, we define a sum operation by using the sum operation in $GF(5^3)$. In this field, the sum is carried out by means of the polynomial sum in the usual fashion with polynomial coefficients reduced by module 5.

Then, for all pair of codons $X_1Y_1Z_1 \in C_{125}$ and $X_2Y_2Z_2 \in C_{125}$, their sum "+" will be:

$$X_1Y_1Z_1 + X_2Y_2Z_2 = \Psi[\Psi^{-1}(X_1Y_1Z_1) + \Psi^{-1}(X_2Y_2Z_2) \bmod 5]$$

As a result, the set of codons ($C_{125}$, +) with operation "+" is an Abelian group and the set ($C_{125}$, +, •) is a field isomorphic to $GF(5^3)$. After that, we can define the product of a codon $XYZ \in C_{125}$ by the element $\alpha_I \in GF(5^3)$. For all $\alpha_I \in GF(5^3)$ and for all $XYZ \in C_{125}$, this operation will be defined as:

$$\alpha_i (XYZ) = \Psi[\alpha_i \Psi^{-1}(XYZ) \bmod 5]$$

This operation is analogous to the multiplication rule of a vector by a scalar. So, ($C_{125}$, +) can be considered a one-dimensional vector space over $GF(5^3)$. The canonical base of this space is the triplet OOA. We shall call this structure the vector space of extended triplets over $GF(5^3)$. Such structure can be extended to the $N$-dimensional sequence space ($S$) consisting of the set of all $125^N$ DNA alignment sequences with $N$ extended triplets. Obviously, this set is isomorphic to the set of all $N$-tuples ($x_1,...,x_N$) where $x_i \in C_{125}$.

Next, set $S$ can be represented by all $N$-tuples ($x_1,...,x_N$) $\in (C_{125})^N$. As a result, the $N$-dimensional vector space of $S$ over $GF(5^3)$ will be the direct sum

$$S = (C_{125})^N = C_{125} \oplus C_{125} \oplus ... \oplus C_{125} \text{ (N times)}$$

The sum and product in $S$ are carried out by components (Redéi, 1967). That is, for all $\alpha \in GF(5^3)$ and for all $s, s' \in S$ we have:

$$s + s' = (s_1, s_2,..., s_N) + (s_1', s_2',..., s_N') = (s_1 + s_1', s_2 + s_2',..., s_N + s_N')$$

$$\alpha s = \alpha (s_1, s_2... s_N) = (\alpha s_1, \alpha s_2... \alpha s_N)$$

TABLE I
ORDERED SET OF EXTENDED TRIPLETS CORRESPONDING TO THE ELEMENTS OF *GF* (5³).

| A | | O | | | A | | | C | | | G | | | U | | | |
|---|---|---|---|---|---|---|---|---|---|---|---|---|---|---|---|---|---|
| | I | II | III | I | II | III | I | II | III | I | II | III | I | II | III | |
| O | 0 | 000 | OOO | 25 | 001 | OAO | 50 | 002 | OCO | 75 | 003 | OGO | 100 | 004 | OUO | O |
| | 1 | 100 | OOA | 26 | 101 | OAA | 51 | 102 | OCA | 76 | 103 | OGA | 101 | 104 | OUA | A |
| | 2 | 200 | OOC | 27 | 201 | OAC | 52 | 202 | OCC | 77 | 203 | OGC | 102 | 204 | OUC | C |
| | 3 | 300 | OOG | 28 | 301 | OAG | 53 | 302 | OCG | 78 | 303 | OGG | 103 | 304 | OUG | G |
| | 4 | 400 | OOU | 29 | 401 | OAU | 54 | 402 | OCU | 79 | 403 | OGU | 104 | 404 | OUU | U |
| A | 5 | 010 | AOO | 30 | 011 | AAO | 55 | 012 | ACO | 80 | 013 | AGO | 105 | 014 | AUO | O |
| | 6 | 110 | AOA | 31 | 111 | AAA | 56 | 112 | ACA | 81 | 113 | AGA | 106 | 114 | AUA | A |
| | 7 | 210 | AOC | 32 | 211 | AAC | 57 | 212 | ACC | 82 | 213 | AGC | 107 | 214 | AUC | C |
| | 8 | 310 | AOG | 33 | 311 | AAG | 58 | 312 | ACG | 83 | 313 | AGG | 108 | 314 | AUG | G |
| | 9 | 410 | AOU | 34 | 411 | AAU | 59 | 412 | ACU | 84 | 413 | AGU | 109 | 414 | AUU | U |
| C | 10 | 020 | COO | 35 | 021 | CAO | 60 | 022 | CCO | 85 | 023 | CGO | 110 | 024 | CUO | O |
| | 11 | 120 | COA | 36 | 121 | CAA | 61 | 122 | CCA | 86 | 123 | CGA | 111 | 124 | CUA | A |
| | 12 | 220 | COC | 37 | 221 | CAC | 62 | 222 | CCC | 87 | 223 | CGC | 112 | 224 | CUC | C |
| | 13 | 320 | COG | 38 | 321 | CAG | 63 | 322 | CCG | 88 | 323 | CGG | 113 | 324 | CUG | G |
| | 14 | 420 | COU | 39 | 421 | CAU | 64 | 422 | CCU | 89 | 423 | CGU | 114 | 424 | CUU | U |
| G | 15 | 030 | GOO | 40 | 031 | GAO | 65 | 032 | GCO | 90 | 033 | GGO | 115 | 034 | GUO | O |
| | 16 | 130 | GOA | 41 | 131 | GAA | 66 | 132 | GCA | 91 | 133 | GGA | 116 | 134 | GUA | A |
| | 17 | 230 | GOC | 42 | 231 | GAC | 67 | 232 | GCC | 92 | 233 | GGC | 117 | 234 | GUC | C |
| | 18 | 330 | GOG | 43 | 331 | GAG | 68 | 332 | GCG | 93 | 333 | GGG | 118 | 334 | GUG | G |
| | 19 | 430 | GOU | 44 | 431 | GAU | 69 | 432 | GCU | 94 | 433 | GGU | 119 | 434 | GUU | U |
| U | 20 | 040 | UOO | 45 | 041 | UAO | 70 | 042 | UCO | 95 | 043 | UGO | 120 | 044 | UUO | O |
| | 21 | 140 | UOA | 46 | 141 | UAA | 71 | 142 | UCA | 96 | 143 | UGA | 121 | 144 | UUA | A |
| | 22 | 240 | UOC | 47 | 241 | UAC | 72 | 242 | UCC | 97 | 243 | UGC | 122 | 244 | UUC | C |
| | 23 | 340 | UOG | 48 | 341 | UAG | 73 | 342 | UCG | 98 | 343 | UGG | 123 | 344 | UUG | G |
| | 24 | 440 | UOU | 49 | 441 | UAU | 74 | 442 | UCU | 99 | 443 | UGU | 124 | 444 | UUU | U |

[a] In this table it is possible to see the bijection between the triplet set and the set of 3-tuples in $(Z_5)^3$, which are also the coefficients of the polynomials in the *GF* (5³). The corresponding integer number of every 3-tuples is also shown. I. Triplet index number. II. Polynomial coefficients. III. Extended triplets

Next, it can be proved that $(S, +)$ is an Abelian group with the *N*-tuple $s_e$ = (OOO, OOO…OOO) as its neutral element. The canonical base of this space is the set of vectors:

$$e_1 = (OOA, OOO, \ldots, OOO), e_2 = (OOO, OOA, \ldots, OOO), \ldots, e_N = (OOO, OOO, \ldots, OOA)$$

As a result, every sequence $s \in S$ has a unique representation:

$$s = \alpha_1 e_1 + \alpha_2 e_{1+\ldots} + \alpha_N e_N \; (\alpha_I \in GF(5^3))$$

It is usually said that the *N*-tuple $(\alpha_1, \alpha_2, \ldots, \alpha_N)$ is the coordinate representation of *s* in the canonical bases $\{e_i \in C_{125}, i=1,2,\ldots,N\}$ of *S*.

### III. RESULTS AND DISCUSSION

Evidently, the Galois field of codons is not unique. Actually, we have obtained forty isomorphic Galois fields, each one with the product operation defined from one of the forty irreducible polynomials. It is convenient, however, to choose a most biologically significant Galois field.

The most attractive irreducible polynomials are the primitive polynomials. If $\alpha$ is a root of a primitive polynomial then its powers $\alpha^n$ ($n = 1,\ldots, 124$) are the elements of the multiplicative group of $GF(125)$, i.e. $\alpha$ is a group generator. As it was shown in [SAN 05], a product operation in a Galois field generated by a primitive polynomial is carried out in a very simple way (see Table II). Just twenty of the forty irreducible polynomials are primitives. In [SAN 05a] the sum operation is a manner to consecutively obtain all codons from the codon AAC in such a way that the genetic code will represent a non-dimensional code scale of amino acid interaction energy in proteins. Here, in order to consecutively obtain all codons from the codon AAC we choose those primitive polynomials with root $\alpha = 2 + x + x^2$ –corresponding to codon AAC. Only primitive polynomial $g(x) = 2 + 3 x^2 + x^3$ has this root, in this way the product operation is unique.

Notice that, in the vector space $S$ are represented all $125^N$ possible DNA alignment sequences of length $N$. Here the "classical gap", produced by alignment algorithms corresponds to the neutral element "O". The neutral element appears from algebraic operations with codons. For instance, in the additive group $(C_{125}, +)$, the inverse of codons $X_1AX_3$ coding to hydrophilic amino acids are the codons $(-X_1)U(-X_3)$ that in turn code to hydrophobic amino acids. The sum of a $X_1AX_3$ codon to a $X_1UX_3$ codon produces a $X_1OX_3$ codon. Then, this sum introduces, at least, one base deletion in the obtained extended triplet. In general, indel mutations found in the molecular evolution process can be described by means of algebraic operations in $(C_{125}, +, \bullet)$, i.e. any deletion or insertion presented in any mutant DNA sequence are described by means of algebraic transformations of the corresponding wild type gene.

## A. Transformations of the DNA Extended Sequences

Gene mutations can be considered as linear transformations of the wild type gene in the $N$-dimensional vector space of DNA sequences. These linear transformations are endomorphisms and automorphisms. In particular, there are some remarkable automorphisms. Automorphisms are one-one transformations on the group $(C_{125})^N$, such that:

$$f(a \cdot (\alpha+\beta)) = a f(\alpha) + a f(\beta) \text{ for all extended DNA sequences } \alpha \text{ and } \beta \text{ in } (C_{125})^N \text{ and } a \in GF(124)$$

That is, automorphisms forecast mutation reversions, and if the molecular evolution process went by through automorphisms then, the observed current DNA sequences would not depend on the mutational pathway followed by the ancestral DNA sequences. In addition, the set of all automorphisms is a group.

For every endomorphism (or automorphism) $f: (C_{125})^N \to (C_{125})^N$, there is a $N \times N$ matrix:

$$A = \begin{pmatrix} a_{11} & \ldots & a_{1N} \\ . & . & . \\ a_{N1} & \ldots & a_{NN} \end{pmatrix}$$

with rows that are the image vectors $f(e_i)$, $i=1,2,\ldots N$. This matrix will be called the representing matrix of the endomorphism $f$ with respect to the canonical base $e_i$ $\{i=1,2\ldots,N\}$.

As in [SAN 05], single point mutations can be considered local endomorphisms. An endomorphism $f: S \to S$ will be called local endomorphism if there exists $k \in \{1, 2,\ldots, N\}$ and $a_{ik} \in GF(125)$ ($i = 1, 2,\ldots,N$) such that:

$$f(e_i) = a_{ik}e_k + e_i, \text{ for } i \neq k,$$

and
$$f(e_k) = a_{kk}e_k$$
This means that:
$$f(x_1, x_2, ...x_n) = (x_1, x_2, ...\sum_{i=1}^{n} x_i a_{ik}, ...x_n)$$

It is evident that a local endomorphism will be a local automorphism if, and only if, the element $a_{kk}$ is different from cero. The local endomorphism $f$ will be considered diagonal if $f(e_k) = (0,...,a_{kk},...,0) = a_{kk}e_k$ and $f(e_i) = e_i$, for $i \neq k$. This means that:

$$f(x_1, x_2, ...x_N) = (x_1, x_2, ...a_{kk} x_k, ...x_N)$$

The previous concepts allow us to present the following theorem:

**Theorem 1**. For every single point mutation changing the codon $\alpha_i$ of the wild type gene $\alpha = (\alpha_1, \alpha_2, ..., \alpha_i, ..., \alpha_N)$ ($\alpha$ different from the null vector) by the codon $\beta_i$ of the mutant gene $\beta = (\alpha_1, \alpha_2, ..., \beta_i, ..., \alpha_N)$, there is:
  i.  At least a local endomorphism $f$ such that $f(\alpha) = \beta$.
  ii. At least a local automorphism $f$ such that $f(\alpha) = \beta$.
  iii. A unique diagonal automorphism $f$ such that $f(\alpha) = \beta$ if, and only if, the codons $\alpha_i$ and $\beta_i$ of the wild type and mutant genes, respectively, are different of GGG.

Proof: Since genes are included in the vector space over a Galois Field, this prove is similar to those reported in [SAN 05]. □

According to the last theorem, any mutation point presented in the vector space $(C_{125})^N$ of all DNA alignment sequences of length $N$ sequences are described by means of automorphisms of the corresponding wild type gene. Specifically, the most frequent mutation can be described by means of diagonal automorphisms [see SAN 05]. We can consider, for example, the sequence α=UAUAUGAGUGAC. Let us suppose that, with successive mutations, this sequence become the sequence β = UGUAUAAGUOAG. According to Table 1 these sequences correspond in the vector space $(C_{125})^4$ to vectors $\alpha$ = (f [24], f [108], f [84], f [42]) and $\beta$ = (f [99], f [106], f [84], f [28]). Hence, according to the Theorem, there exists a diagonal endomorphism $f$, so that $\beta = f(\alpha)$. Our Galois field is generated by the primitive polynomial $g(x) = 2 + 3 x^2 + x^3$. In this field the root $\alpha = 2 + x + x^2$ –corresponding to codon AAC– is a generator of the multiplicative group. Next, by means of Table II we can compute:

$$(f[99], f[106], f[84], f[28]) = (f[24], f[108], f[84], f[42]) \begin{pmatrix} f[55] & 0 & 0 & 0 \\ 0 & f[61] & 0 & 0 \\ 0 & 0 & f[1] & 0 \\ 0 & 0 & 0 & f[29] \end{pmatrix}$$

On the other hand, mutations can be considered translations of the wild type gene in the $N$-dimensional vector space of the DNA extended sequences. In the Abelian group $(C_{125}, +)$, for two extended triplets $a, b \in (C_{125}, +)$, equation $a + x = b$ always has a solution. Then, for all pair of alignment sequences $\alpha, \beta \in (C_{125}, +)^N$ there is always a sequence $\kappa \in (C_{125}, +)^N$ so that $\alpha + \kappa = \beta$. That is, there exists translation $T : \alpha \to \beta$. We shall represent translation $T_k$ with constant $k$ acting on triplet $x$ as:
$$T_k(x) = x + k$$
Next, given applications: $W \xrightarrow{f} X \xrightarrow{g} Y$, the composition $g \circ f : W \to Y$ of translations $g$ and $f$ is defined by $(g \circ f)(x) = g(f(x))$. It is not difficult to see that the set of all translations with composition operations is a group $G$.

## B. Stabilizer subgroup of the wild type conserved regions

It is well known that in a wild type ORF, normally, not all codon sequences are susceptible to experimental mutations. Usually, conserved variables and hypervariables regions are found in genes. A typical case is the antibody where heavy chain variable domain ($V_H$) and a light chain variable domain ($V_L$) are found. Within $V_L$ and $V_H$ there are "hot spots" of variability. These hot spots of variability were termed hypervariable regions. The hypervariable regions of the heavy and light chains together form the antigen binding site of the immunoglobulin molecule. Next, let $P$ be the subset of mutant DNA sequences conserving the same regions from a wild type DNA coding sequence $\alpha_0 \in (C_{125})^N$. Then, according to the group theory [RED 67], the set $St(\alpha_0)$ of automorphisms $f \in G$ that preserves these regions is a subgroup of $G$, that is:

$$St(\alpha_0) = \{f \in G, \text{ such that: } f(\alpha_0) = \beta \in P\} \subset G$$

This subgroup could be called the stabilizer subgroup in $G$ of the conserved regions of wild type $\alpha_0$. Notice that the stabilizer subgroup $St(\alpha_0)$ is connected with the homologous recombination that involves a reciprocal exchange of DNA sequences −e.g. between two chromosomes that carry the same genetic loci. The homologous recombination algebraically corresponds to the action of two automorphism pairs that could be included in the $St(\alpha_0)$ (see Fig. 1). For instance, the pair $f$ and $f^{-1}$ acts over the homologous strands $\alpha$ and $\beta$ to produce the homologous reciprocal recombinants $f(\alpha)$ and $f^{-1}(\beta)$. Likewise, the pair $g$ and $g^{-1}$ acts over the homologous strands $\alpha'$ and $\beta'$ to produce the homologous reciprocal recombinants $g(\alpha')$ and $g^{-1}(\beta')$. As a result, two reciprocal recombinant DNA sequences are generated. In particular, if homologous recombination results in an exact exchange of genetic information, then the automorphism pairs are diagonal automorphisms. Since evolution could not happen without genetic recombination, this algebraic description is biologically relevant. If it were not possible to exchange material between (homologous) chromosomes, the content of each individual chromosome would be irretrievably fixed in its particular alleles. When mutations occurred, it would not be possible to separate favourable and unfavourable changes [LEW 04]. Hence, the study of the automorphism subgroup involved in this transformation −the homologous recombination− could reveal new rules of molecular evolution process so far unknown.

## C. Finite Abelian group of DNA sequences

Now, we would like to analyse some subset of DNA alignment sequences with length $N$. By means of multiple sequence alignments it is possible to find in the DNA genomic sequences small subregions in which there are not introduced gaps. For instance, if we observe multiple sequence alignments of open reading frames (ORF) from gene super families we can detect small blocks of ungaped aligned sequences from different genes. We shall call these small blocks of codon sequences "building blocks" (see Fig 2). Theoretically, a building block will be a set of aligned sequences $X_1 X_2 \ldots X_N$ where $X_i \in \{A, C, G, U\}$ ($i = 1 \ldots N$) with some evolutionary relationship between them. The building blocks with length $N$ can be described by vector spaces over $GF(64)$; in particular, these can be described by the Abelian group $((C_{64})^N, +)$ of the $N$-dimensional vector space of DNA sequences. Whereas regions with gaps can be described by means of the group $((C_{125})^N, +)$.

Notice that groups $(C_{64}, +)$ and $(C_{125}, +)$ are isomorphic to the $p$-groups ($p$ a prime number) $(Z_2)^3$ (a 2-group) and $(Z_5)^3$ (a 5-group) respectively. It is well known that every Abelian group can be written as a direct sum of $p$-groups [DUB 63]. Actually, in the set of all alignment sequences with length $N$ we can define several finite Abelian groups over the subsets of all $2^{m_1+m_2+\ldots+m_p} 5^{n_1+n_2+\ldots+n_q}$ possible alignment sequences ($N = n_1+\ldots+n_p+m_1+\ldots+m_q$). An example of these is showed in Fig 1. We shall call these group "alignment groups".

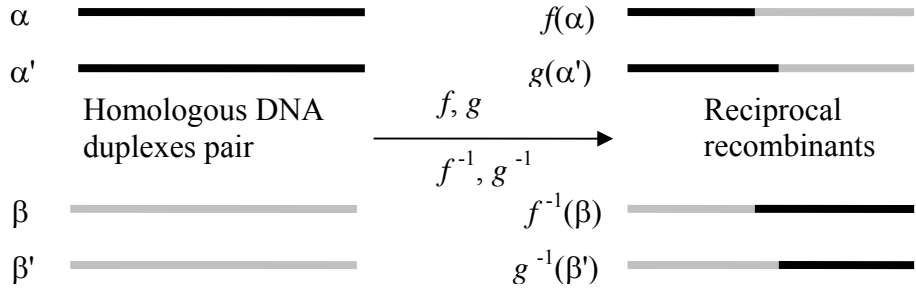

Fig. 1 The homologous (generalized) recombination between two homologous DNA duplexes algebraically corresponds to the action of two automorphism pairs over two paired DNA duplexes. The two automorphism pairs express a reciprocal exchange of DNA sequences and could be included in the subgroup of automorphism $St\,(\alpha_0)$.

TABLE II.
LOGARITHM TABLE OF THE ELEMENTS OF THE $GF(5^3)$ GENERATED BY THE PRIMITIVE POLYNOMIAL $g(x) = 2 + 3\,x^2 + x^3$.

| Element | f[1] | f[2] | f[3] | f[4] | f[5] | f[6] | f[7] | f[8] | f[9] | f[10] | f[11] | f[12] |
|---|---|---|---|---|---|---|---|---|---|---|---|---|
| $n$ [1] | 0 | 93 | 31 | 62 | 25 | 4 | 42 | 105 | 102 | 118 | 74 | 97 |
| Element | f[13] | f[14] | f[15] | f[16] | f[17] | f[18] | f[19] | f[20] | f[21] | f[22] | f[23] | f[24] |
| $n$ | 71 | 11 | 56 | 73 | 9 | 35 | 12 | 87 | 40 | 43 | 104 | 66 |
| Element | f[25] | f[26] | f[27] | f[28] | f[29] | f[30] | f[31] | f[32] | f[33] | f[34] | f[35] | f[36] |
| $n$ | 50 | 23 | 30 | 28 | 106 | 29 | 14 | 1 | 20 | 86 | 67 | 8 |
| Element | f[37] | f[38] | f[39] | f[40] | f[41] | f[42] | f[43] | f[44] | f[45] | f[46] | f[47] | f[48] |
| $n$ | 83 | 120 | 38 | 6 | 80 | 46 | 10 | 79 | 3 | 119 | 95 | 109 |
| Element | f[49] | f[50] | f[51] | f[52] | f[53] | f[54] | f[55] | f[56] | f[57] | f[58] | f[59] | f[60] |
| $n$ | 84 | 19 | 121 | 116 | 75 | 123 | 99 | 103 | 49 | 48 | 15 | 122 |
| Element | f[61] | f[62] | f[63] | f[64] | f[65] | f[66] | f[67] | f[68] | f[69] | f[70] | f[71] | f[72] |
| $n$ | 113 | 107 | 55 | 94 | 96 | 78 | 88 | 53 | 64 | 36 | 89 | 101 |
| Element | f[73] | f[74] | f[75] | f[76] | f[77] | f[78] | f[79] | f[80] | f[81] | f[82] | f[83] | f[84] |
| $n$ | 7 | 52 | 81 | 61 | 13 | 54 | 59 | 98 | 114 | 69 | 39 | 27 |
| Element | f[85] | f[86] | f[87] | f[88] | f[89] | f[90] | f[91] | f[92] | f[93] | f[94] | f[95] | f[96] |
| $n$ | 34 | 2 | 115 | 26 | 16 | 60 | 32 | 117 | 45 | 51 | 37 | 77 |
| Element | f[97] | f[98] | f[99] | f[100] | f[101] | f[102] | f[103] | f[104] | f[105] | f[106] | f[107] | f[108] |
| $n$ | 110 | 111 | 41 | 112 | 44 | 90 | 92 | 85 | 65 | 22 | 47 | 33 |
| Element | f[109] | f[110] | f[111] | f[112] | f[113] | f[114] | f[115] | f[116] | f[117] | f[118] | f[119] | f[120] |
| $n$ | 57 | 68 | 17 | 72 | 108 | 18 | 5 | 100 | 58 | 21 | 70 | 91 |
| Element | f[121] | f[122] | f[123] | f[124] | | | | | | | | |
| $n$ | 24 | 82 | 63 | 76 | | | | | | | | |

Here, codon AAC corresponds to the primitive root $\alpha = 2 + x + x^2$, i.e. $f[s] = (2 + x + x^2)^n \bmod g(x)$ and $n$ = logarithm base $\alpha$ of $f[s]$ = $\log_\alpha f[s]$. The properties of this logarithm function are alike to the classical definition in arithmetic:
  i. $\log_\alpha (f[x]*f[y]) = (\log_\alpha f[x] + \log_\alpha f[y])\ mod\ 124 = (n_x + n_y)\ mod\ 124$
  ii. $\log_\alpha (f[x]/f[y]) = (\log_\alpha f[x] - \log_\alpha f[y])\ mod\ 124 = (n_x - n_y)\ mod\ 124$
  iii. $\log_\alpha f[x]^m = m\,\log_\alpha f[x]\ mod\ 124$

Since the canonical decomposition of an Abelian group $G$ into $p$-groups is a unique and safe isomorphism [DUB 63], it is possible to characterize alignment groups for a fixed sequence length $N$. That is to say, two alignment groups can have different $p$-group decompositions and simultaneously be isomorphic by holding the same canonical decomposition into $p$-groups. This algebraic description biologically suggests that the same biological architectural principium underlies the alignment groups with the same canonical decomposition into $p$-groups. Here the basic construction materials come from building blocks. It could also correspond to the fact that in the molecular evolution process the new genetic information frequently comes into being from the rearrangements of existing genetic material in the chromosomes.

If a finite group $G$ is written as a direct sum $G = G_1 \oplus G_2 \oplus \ldots \oplus G_s$, then endomorphism ring End($G$) is isomorphic to the ring matrices $(A_{ij})$, where $A_{ij} \in \text{Hom}(G_i, G_j)$, with the usual matrix operations. In our case the endomorphism that transform the DNA alignment sequence $\alpha$ into $\beta$ ($\alpha, \beta \in G$) is represented by a matrix with only non-cero elements in the principal diagonal. These diagonal elements are sub-matrices $A_{i\ i} \in$ End($G_i$) (o $A_{ii} \in$ Aut($G_i$)).

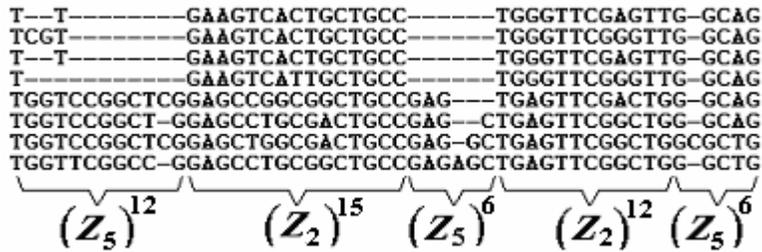

Fig. 2 An example of alignment group $S = (Z_5)^{12} \oplus (Z_2)^{15} \oplus (Z_5)^6 \oplus (Z_2)^{12} \oplus (Z_5)^6$. Building blocks correspond to ungaped sub-sequences (power of $Z_2$).

## IV. CONCLUSIONS

In this paper the extend triplet set with elements $X_1X_2X_3$ where $X_i \in \{O, A, C, G, U\}$ is the starting point to analyze deletions and mutations in DNA sequences. Taking into account the order in the set $\{O, A, C, G, U\}$ and the biological importance of base positions in the codon, it is possible to establish a bijection between the extended triplet set and the Galois field GF($5^3$). This bijection allows us to define the Galois field of the extended triplet set. Over this new field, a new $N$-dimensional vector space is defined in the set of all possible DNA alignment sequences where gene mutations can be considered linear transformations or translations of the wild type gene.

For every single point mutation in the wild type gene there is at least an automorphism that transforms the wild type in the mutant gene. So, automorphisms group could be a useful tool to study the mutational pathway followed by genes in the $N$-dimensional vector space of all possible DNA alignment sequences.

Besides this, the set $St(\alpha_0)$ of automorphisms that conserve the same regions from a wild type DNA coding sequence $\alpha_0 \in (C_{125})^N$ is a subgroup connected with the homologous recombination that involves a reciprocal exchange of DNA sequences −e.g. between two chromosomes that carry the same genetic loci. The homologous recombination algebraically corresponds to the action of two automorphism pairs that could be included in the $St(\alpha_0)$

By means of multiple sequence alignments it is possible to define several finite Abelian groups −alignment groups− over the subsets of all $2^{m_1+m_2+\ldots+m_p} 5^{n_1+n_2+\ldots+n_q}$ possible alignment sequences ($N = n_1+\ldots+n_p+m_1+\ldots+m_q$). Two alignment groups can have different $p$-group decompositions and simultaneously be isomorphic holding the same canonical decomposition into $p$-groups. For alignment groups with the same canonical *p*-group decompositions could underlie the same biological architectural principium.

ACKNOWLEDGMENTS

This research was supported within the framework of a VLIR-IUS Collaboration Programme.